\documentclass[a4paper]{article}

\usepackage{INTERSPEECH2022}
\usepackage{url}
\usepackage{hyperref}

\title{RetrieverTTS: Modeling Decomposed Factors for Text-Based Speech Insertion}
\name{Dacheng Yin$^{1*}$\thanks{*Work done during internship at Microsoft Research Asia.}, Chuanxin Tang$^2$, Yanqing Liu$^3$, Xiaoqiang Wang$^3$, Zhiyuan Zhao $^2$, Yucheng Zhao$^1$, Zhiwei Xiong$^1$, Sheng Zhao$^3$, Chong Luo$^2$}
\address{
  $^1$University of Science and Technology of China, Hefei, China\\
  $^2$Microsoft Research Asia, Beijing China \\
  $^3$Microsoft Azure Speech, Beijing, China}
\email{\{ydc,lnc\}@mail.ustc.edu.cn, zwxiong@ustc.edu.cn, \{chutan,yanqliu,xiaoqwa,zhiyzh,szhao,cluo\}@microsoft.com}

\begin{document}

\maketitle

\begin{abstract}
This paper proposes a new ``decompose-and-edit" paradigm for the text-based speech insertion task that facilitates arbitrary-length speech insertion and even full sentence generation. 
In the proposed paradigm, global and local factors in speech are explicitly decomposed and separately manipulated to achieve high speaker similarity and continuous prosody.
Specifically, we proposed to represent the global factors by multiple tokens, which are extracted by cross-attention operation and then injected back by link-attention operation.  Due to the rich representation of global factors, we manage to achieve high speaker similarity in a zero-shot manner.
In addition, we introduce a prosody smoothing task to make the local prosody factor context-aware and therefore achieve satisfactory prosody continuity. 
We further achieve high voice quality with an adversarial training stage. 
In the subjective test, our method achieves state-of-the-art performance in both naturalness and similarity. 

Audio samples can be found at \url{https://ydcustc.github.io/retrieverTTS-demo/}.
\end{abstract}

\section{Introduction}

Nowadays, the demand of video production is growing rapidly. Emerging technology \cite{discript.org,fried2019text,jin2017voco} provides efficient automatic tools for text-based speech or video editing to improve the video producers' efficiency. Such systems support cutting, copying, pasting operations, and even inserting synthesized speech or video segments based on text transcriptions. Among these operations, insertion is the most challenging one and still an open problem. For speech, it requires not only high voice quality, but also high voice similarity and continuous prosody. Our work aims at achieving the above three goals at once. 

Ideally, a text-based speech insertion method should solve a zero-shot speaker-context-conditioned speech synthesis task. Given any utterance and its transcript with the inserted words, the system needs to predict an inserted speech segment that should satisfy temporal smoothness including both speaker consistency and context continuity. Two existing works \cite{chuanxin2021speechinsert, borsos2022speechpainter} are directly aimed at this task.
In \cite{chuanxin2021speechinsert}, the encoded text sequence is inserted into the encoded mel sequence, and then a decoder is applied to implicitly model the temporal smoothness between the insertion and the context, and predict the desired mel-spectrum.
In \cite{borsos2022speechpainter}, Perceiver IO \cite{perceiverio} is directly adopted. Both the temporal smoothness and text-spectrum alignment are achieved implicitly, via Perceiver IO's ``process" and ``decode" stage. 
However, both methods are limited to edit very short speech segments (one word or within a second), due to the difficulty in directly estimating fine-grained mel-spectrogram.
Moreover, \cite{borsos2022speechpainter} uses a large network to solve the problem and the convergence is slow (3.1M steps, 256 batch-size).

Another category of research, speaker conditioned speech generation, is also highly related to our work. It aims at generating speech of the desired speaker's timbre and style given any text as input. A text-based speech insertion system can be derived from such methods by inserting the generated segment into the appropriate position of the original waveform. In this category, two lines of existing research are separately developed. In the first line, methods \cite{adaspeech1, adaspeech2, adaspeech3, metavoice} achieve satisfactory voice quality and similarity by finetuning the pre-trained model on the reference voice before inference. However, such methods require more than one utterance-text pair and hundreds of finetuning steps, which is not applicable in our scenario. In the other line of research, the goal is to learn sufficiently good representations for timbre and style, so that the model achieves zero-shot adaptation without finetuning. The representative methods \cite{tacotron_transfer, metastylespeech} use only a single speaker vector, which is not enough to guarantee a complete modeling of both timbre and prosody. Therefore, the speaker similarity is not sufficiently satisfactory. Moreover, a clear drawback of the derived system is that the generated prosody often mismatches the context.
\begin{figure*}[t]
\vspace{-1.5em}
\centering
\includegraphics[width=\textwidth]{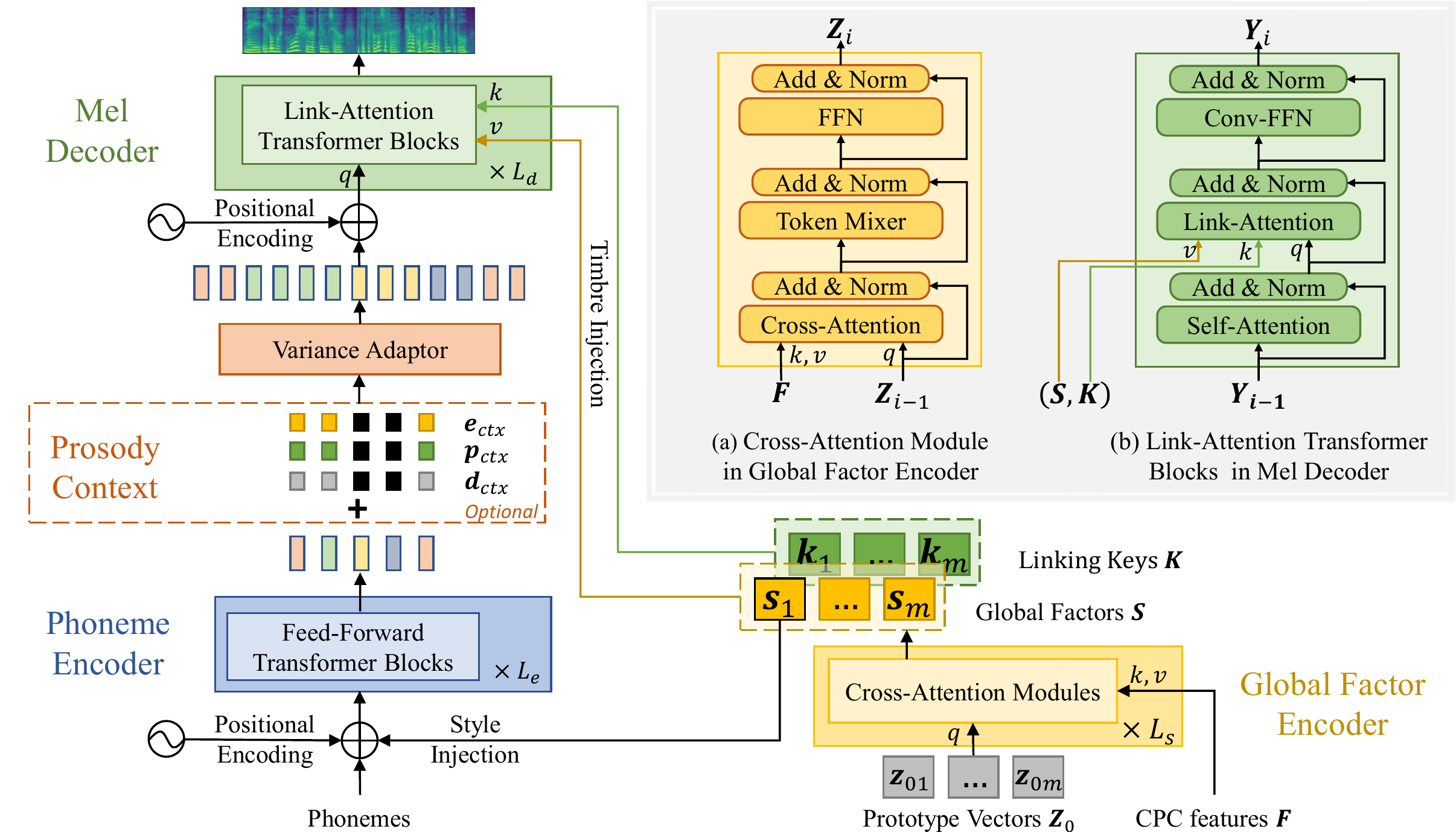}
\vspace{-1.5em}
\caption{\looseness -1 Model architecture.\label{fig:arch}} 
\vspace{-1.5em}
\end{figure*}

A third category of related works is zero-shot voice conversion  \cite{AdaINVC,autovc,FragmentVC,S2VC,yin2022Retriever}. Such methods achieve good disentanglement between speaker and content information. Notably, Retriever \cite{yin2022Retriever} achieves the state-of-the-art result by using a set of style tokens other than a single speaker vector to model speaker information, and the proposed link attention achieves fine-grained speaker information injection. Such voice conversion systems can be used together with a single-speaker TTS model, converting the insertion speech segment generated by the TTS model into any other speaker's voice. However, such pipeline has large redundancy since the voice conversion model re-extracts content from the speech signal even though the inserted content is already specified by the user. Moreover, stacking multiple systems brings the risk of error accumulation.

To solve the speech insertion problem while avoiding the drawbacks of the above methods, we need to analyze the necessary factors for an ideal insertion. We notice that the factors can be divided into global and local factors. The local factor, like prosody, is dependent on the context, since it is highly correlated to the semantic meaning and the sequential information in the utterance. Conversely, global factors like timbre and style are not necessarily dependent on the context. They are consistent among the sentences spoken by the same person, therefore should be given the freedom to be extracted from any other utterances of the same person saying totally different content.

Motivated by this, we decide to build a speech insertion system in a ``decompose-and-edit" paradigm that accurately decomposes global and local factors and manipulates them separately. In general, the speech signal can be divided into four factors: text, prosody, timbre, and style. Here, we characterize prosody as duration, pitch, and energy at the phoneme level. As such, text and prosody are categorized as local factors, while timbre and style are global factors. When new speech is inserted, its text component is specified by the user, while prosody is inpainted according to prosody context, style, and the inserted text. In contrast, the global factors just remain unchanged. This paradigm
isolates global factors from contextual information while keeping local factors context-aware. As such, both full sentence generation and long insertion are supported.

Under this paradigm, we need to achieve the following three key technical goals.
Firstly, it is required to achieve high speaker similarity through precise global factor extraction and injection. Inspired by Retriever \cite{yin2022Retriever}, we use cross-attention modules to extract a set of global factor tokens to model global factors as complete as possible and leverage link-attention in the Mel-decoder to achieve as thorough as possible global factor injection.
Secondly, to explicitly guide our model to generate prosody according to the prosody context, we introduce a  prosody-smoothing task during training. 
Finally, the basic requirement of a speech synthesis system is realistic voice quality. Therefore, we apply adversarial training for this purpose.

In a nutshell, our contributions are the following four folds:
(1) Build a text-based speech editing system by separately controlling global and local factors, enabling powerful speech editing operations including any-length insertion and full sentence generation. (2) With a powerful global factors extraction and injection method, we achieve state-of-the-art speaker similarity. (3) By introducing a prosody smoothing task, we achieve highly continuous prosody for text-based speech insertion. (4) Subjective tests show that our system yields nearly indistinguishable insertion results, and outperforms other competitive methods.

\section{Method}
\subsection{Model Architecture}
To explicitly model local and global factors and control them separately, both factors need explicit representation. For local factors, the non-AR TTS models like FastPitch \cite{fastpitch} satisfies our requirement, since it learns both text and prosody embeddings. Therefore, we develop our speech synthesis procedure from that proposed in FastPitch. For global factors, we apply the state-of-the-art method \cite{yin2022Retriever} to meet our demands. 

As shown in Fig.\ref{fig:arch}, our model consists of four major parts: phoneme encoder, variance adaptor, mel decoder, and global factor encoder. The phoneme encoder encodes phoneme sequence into latent space with a stack of $L_e=6$ feed-forward transformer blocks. Then the variance adaptor predicts duration, pitch, and energy according to the encoded phonemes together with optional prosody context $\{\bm{e}_{ctx}, \bm{p}_{ctx}, \bm{d}_{ctx}\}$. In the variance adaptor, pitch and energy embeddings are added to the encoded phonemes, and then the length regulator (LR) duplicates each phoneme-level embedding vector according to the phoneme duration. Finally, the mel decoder decodes the length-regulated sequence into a mel-spectrum with a stack of $L_d=6$ link-attention transformer blocks.

During the above generative process, global factors are represented by $m$ tokens $\bm{S} = \{\bm{s}_1, \bm{s}_2, ..., \bm{s}_m\}$ and injected into both the encoder and the decoder. For style injection, the first token $\bm{s_1}$ is added to each input token of the phoneme encoder. For timbre injection, all the tokens are injected into the mel decoder. To achieve this, each link-attention transformer block is equipped with a link-attention \cite{yin2022Retriever} in between self-attention and Conv-FFN, as shown in Fig.\ref{fig:arch}  (b). The link-attention is a multi-head attention which takes the self-attention output as query, the global factor tokens $\bm{S}$ as value, and the linking keys $\bm{K}$ as key, where $\bm{K} = \{\bm{k}_1, \bm{k}_2, ..., \bm{k}_m\}$ is a set of learnable vectors paired with the $m$ global factor tokens. This mechanism allows fine-grained timbre injection, because each decoder frame calculates its own weighted sum of the $m$ global factor tokens as its own timbre modulation. 

For global factor encoder, we use a stack of $L_s=3$ cross-attention modules to query the information from the CPC \cite{modifiedcpc} feature $\bm{F}$ that corresponds to the reference speech waveform. A set of learnable prototype vectors $\bm{Z}_0 = \{\bm{z}_{01}, \bm{z}_{02}, ..., \bm{z}_{0m}\}$ are used as the first query of the cross-attention module stack. The details of each cross-attention module are shown in Fig.\ref{fig:arch} (a). It consists of three sub-modules: cross-attention, token mixer, and an FFN based on MLP. Each of them is equipped with residue connection and layer normalization. The cross-attention module takes the previous-layer output as query, and takes $\bm{F}$ as key and value. The token mixer mixes the input tokens using a learnable $m\times m$ weight matrix. 

\subsection{Training Strategy}
Our model involves two stages of training. The goal of the first stage is to achieve high similarity and prosody continuity, without caring too much about the voice quality. To this end, we slightly modify the standard FastPitch training strategy. The global factors are extracted from the CPC feature of the ground-truth waveform. MSE loss is applied on both predicted mel-spectrum $\hat{\bm{X}}$ and the predicted duration $\hat{\bm{d}}$, pitch $\hat{\bm{p}}$, and energy $\hat{\bm{e}}$. During training, the decoder is conditioned on the ground-truth phoneme-level duration $\bm{d}$, pitch $\bm{p}$, and energy $\bm{e}$. To get the ground-truth duration, we use an online alignment module jointly trained with our model using an alignment loss $L_{align}$ \cite{onealignment}. The phoneme-level ground-truth pitch and energy are calculated by averaging the frame-level pitch and energy within each phoneme's time span. The training loss for stage one is:

\begin{align}
\begin{split}
    L_{stg_1} = &\Vert \bm{X} - \hat{\bm{X}}\Vert_2^2 + \alpha_1\Vert \bm{d} - \hat{\bm{d}}\Vert_2^2 + \alpha_2\Vert \bm{p} - \hat{\bm{p}}\Vert_2^2 + \\
    &\alpha_3\Vert \bm{e} - \hat{\bm{e}}\Vert_2^2 + \alpha_4 L_{align}(\bm{t}, \bm{X}),
\end{split}
\end{align}
where $\alpha_i$ is the hyper-parameter for $i\in\{1, 2, 3, 4\}$; $\bm{t}$ is the input phoneme sequence that corresponds to the ground-truth mel-spectrum $\bm{X}$.

We add a prosody smoothing task during training. In 50\% of the training samples, we add partially masked ground-truth prosody (duration, pitch, energy) embeddings $\bm{d}_{ctx}, \bm{p}_{ctx}, \bm{e}_{ctx}$ on the encoded phonemes before feeding them into the variance adaptor. The masked embeddings are filled with zero, and the masked span is uniformly sampled from one to three words. 

The second training stage aims to address the spectrum over-smoothing problem introduced by MSE loss. To this end, MelGAN-like adversarial training is employed. We use hinge loss $L_D$ to train our discriminator, and add an additional feature loss term $L_{feat}$ to train our model. 

\begin{align}
\begin{split}
  &L_D = \text{max}(0, 1-D(\bm{X})) + \text{max}(0, 1+D(\hat{\bm{X}})), \\
  &L_{feat} = \frac{1}{N}\sum_{l=0}^{N}{\frac{1}{d_l}\Vert D_l(\bm{X})- D_l(\hat{\bm{X}})\Vert_1}, 
\end{split}
\end{align}
where $N$ is the number of layers in discriminator, $D_l$ represents the $l$-th layer feature of discriminator, and $d_l$ is the dimension of $D_l$. For discriminator architecture, we follow the design in \cite{borsos2022speechpainter}. The model's second stage training loss is a weighted combination of stage 1 loss and feature loss. Here, we choose $\lambda=10$.
\begin{equation}
    L_{stg_2} = L_{stg_1} + \lambda L_{feat}
\end{equation}

\subsection{Inference}
Our system can work under two inference modes: full sentence generation and words insertion. 

In full sentence generation mode, our system works like the original FastPitch model except that the global factor tokens are extracted from a reference speech, and the text encoder and mel decoder are conditioned on the extracted global factor tokens.

In words insertion mode, the global factor tokens and the original prosody are extracted from the original speech. The words are inserted into text sequence to get the new phoneme sequence. The original prosody is inserted with zero embeddings in the insertion region to serve as prosody context. Given global factors tokens, new phoneme sequence, and prosody context as input, the mel-spectrum of the entire sentence is generated by our model, but only the segment to be inserted is converted into waveform by the vocoder. Finally, the waveform is inserted to the corresponding position of the original waveform.

\section{Experiments}
\subsection{Experimental Setup}
We conduct our experiments on LibriTTS \cite{libritts} dataset. The model is trained on the train-clean-360 split which contains 191 hours of speech spoken by 430 female and 474 male speakers.
We use a publicly-available \footnote{https://github.com/kan-bayashi/ParallelWaveGAN} Hifi-GAN \cite{hifigan} vocoder pre-trained on the same training set to generate waveform from mel-spectrum.

For global factor encoder, $m = 60$, the cross-attention module has 192 channels, and its FFN hidden size is set to 512. $\bm{F}$ is extracted by the pre-trained model from s3prl \cite{s3prl} library. The output is projected to 384 channels by a linear layer to fit the channels to that of the phoneme encoder and mel encoder.

For phoneme encoder and mel decoder, all the intermediate features have 384 channels except that the hidden layers in Conv-FFN have 1536 channels. The Conv-FFN consists of two 1-d conv layers with kernel size 3. ReLU is used as the activation function.
The predictors for duration, pitch, and energy share the same architecture. They are composed of three 1-d conv layers with kernel size 3. The output channel numbers of the three layers are 256,256,1, respectively. Each conv layer is followed by ReLU, Layer Norm, and Dropout layers. Dropout is set to 0.1.

In both training stages, we use a batch size of 256, weight decay of 1e-6, and LAMB \cite{lamb} optimizer with $\beta_1=0.9, \beta_2=0.98$. For the first training stage, we apply polynomial learning-rate schedule with learning rate of 0.1, power of 0.5, 1k warm-up steps, and 80k total training steps. In the second stage, we use a fixed learning rate and train for 20k steps. The learning rates for the discriminator and our model are 0.00005 and 0.0001, respectively. Our discriminator receives 32-frame mel-spectrum chunks as input. The loss weights $\alpha_1,\alpha_2,\alpha_3,$ and $\alpha_4$ are set to 0.1, 0.1, 0.1, and 1, respectively.
\subsection{Evaluation method}
We rely on Mean Opinion Score (MOS) evaluations based on subjective listening tests with rating scores from 1 to 5. The tests are along two dimensions: naturalness and similarity, both of which are tested with out-of-training-set speakers, with the scores denoted as MOS and SMOS, respectively. We report the scores with 95\% confidence interval.

For speech naturalness, we randomly sample 25 different speakers from LibriTTS test-clean set. For each speaker, one sentence is randomly selected for testing. In each sentence, a randomly selected segment of $k$ consecutive words is removed, and the remaining speech is used as the context for the speech insertion task. We then let the tested systems recover the removed segments according to the full text and the remaining speech waveform. 

For speech similarity, we randomly sample 50 target utterances from the 32 speakers in LibriTTS test-clean set for testing. Each is paired with a random reference utterance from the same speaker.
The tested system is required to generate a whole sentence, given the text of the target speech and using the reference speech as timbre and style condition.

\begin{table}[t]
\centering 
\vspace{-0.5em}
\caption{Insertion length robustness test. \label{table:insert_len}}
\vspace{-0.5em}
\begin{tabular}{cc}
\toprule
                      & MOS                   \\ \midrule
ground-truth          & 4.01 $\pm$ 0.17       \\
short insert          & 3.97 $\pm$ 0.17       \\
mid insert            & 3.98 $\pm$ 0.17       \\
long insert           & 3.83 $\pm$ 0.16       \\ 
full generation       & 3.85 $\pm$ 0.19       \\ \bottomrule
\end{tabular}
\vspace{-1.5em}

\end{table}

\subsection{Insertion Length Robustness Test}
This test is to show the robustness of our model to the inserted word number $k$. We prepare four test settings: short insertion: $k\in [1, 3]$; mid insertion: $k\in [3, 5]$; long insertion: increasing $k$ until the inserted speech length is longer than 2 seconds; full generation: generate the whole sentence according to the timbre and style extracted from another reference speech spoken by the same speaker. We also calculate the average inserted time span for the first three cases, which are 0.63s, 1.28s, and 2.18s for short, mid, and long insertion, respectively.
The speech naturalness test results are shown in Table \ref{table:insert_len}. We find that the short and mid insertion are almost imperceptible to humans. Though there is reasonable MOS score drop in long insertion and full generation, the model still achieves satisfactory naturalness. This demonstrates our design enables high robustness across various insertion lengths.
\subsection{Ablation Study}
In this test, we analyze the effectiveness of our three key designs: the global factor modeling method, the prosody smoothing training task, and the adversarial training stage. Firstly, a speech insertion naturalness test is conducted on the systems that remove any of these three designs, as shown in Table \ref{table:ablation1}. All the experiments are done in the ``mid insertion" setting. The ground-truth recordings and our full method's speech insertion results are denoted as ``Ground truth" and ``Full method", respectively. In ``-adv" experiment, we remove the second stage training. In ``-prosody-smooth" experiment, we remove the prosody-smoothing task during training, which can be seen as a pure TTS version of our system. In ``- retriever" experiment, we replace the cross-attention based global factor encoder with the speaker encoder in Meta-StyleSpeech \cite{metastylespeech}, and for style injection, the link-attention module is replaced by SALN proposed in \cite{metastylespeech}. We observe that removing any of our key designs will cause a significant naturalness drop. Specifically, we observe robotic artifacts in ``-adv" experiment, unnatural prosody change in ``-prosody-smooth" experiment, and voice similarity issues in ``-retriever" experiment.

\begin{table}[th]
\centering 
\vspace{-0.5em}
\caption{Ablation study on our system. \label{table:ablation1}}
\vspace{-0.5em}
\begin{tabular}{ccc}
\toprule
                     & MOS@mid                     & SMOS            \\ \midrule
Ground truth         & 4.27 $\pm$ 0.21             & 4.45 $\pm$ 0.19 \\
Full method          & \textbf{3.93 $\pm$ 0.23}    & \textbf{3.70 $\pm$ 0.25} \\
- adv                & 3.53 $\pm$ 0.23             & --- \\
- prosody-smooth     & 3.82 $\pm$ 0.22             & --- \\ 
- retriever          & 3.75 $\pm$ 0.22             & 3.60 $\pm$ 0.27 \\ \bottomrule
\end{tabular}
\vspace{-1.5em}

\end{table}

We conduct a similarity test to further confirm the superiority of our global factor extraction and injection method. In Table \ref{table:ablation1}, we observe a significant SMOS drop when replacing our global factor extraction and injection method with that proposed in \cite{metastylespeech}.

\subsection{System Comparison}
In this section, we compare our method with methods belonging to two other paradigms. Specifically, we take \cite{chuanxin2021speechinsert} as a representative of the implicit modeling paradigm, and \cite{metastylespeech} as a representative of using state-of-the-art zero-shot voice adaptative TTS system. 

Notably, the second method has no existing pipeline for speech insertion. For fair comparison, we design a reasonable speech insertion pipeline for it. The TTS system extracts the speaker vector from the original speech and accordingly generates the waveform of the post-insertion text sequence. The Montreal Forced Alignment tool \cite{mfa} is used to get the word-level alignment. The waveform segment of the insertion words is cut from the generated speech and then inserted into the corresponding position of the original speech. 
This pipeline allows us to compare more fairly to other speech insertion systems than generating the inserted words only.

Both naturalness and similarity tests are conducted. Here, we choose the ``long insert" version of the naturalness test. Firstly, our system achieves the highest MOS score. We observe that the implicit modeling method almost fails in the long insertion case, so it is not suitable for the similarity test which requires whole sentence generation. We also find that our system's SMOS score largely outperforms the state-of-the-art zero-shot voice adaptive TTS method, which leverages an explicit speaker-related meta-learning objective. This indicates that our system is powerful enough to handle the personalized TTS task, even though our design does not fully aim at this goal.

\begin{table}[th]
\centering 
\vspace{-0.5em}
\caption{System comparison. \label{table:cpr}}
\vspace{-0.5em}
\begin{tabular}{ccc}
\toprule
                     & MOS@long              & SMOS            \\ \midrule
Ground-Truth         & 4.20 $\pm$ 0.14       & 4.45 $\pm$ 0.19 \\ 
C.Tang et al. \cite{chuanxin2021speechinsert}             & 2.73 $\pm$ 0.25       & ---             \\ 
Meta-StyleSpeech \cite{metastylespeech}     & 3.88 $\pm$ 0.20       & 2.91 $\pm$ 0.31 \\
RetrieverTTS (Ours)                 & \textbf{3.95 $\pm$ 0.17}       & \textbf{3.70 $\pm$ 0.25} \\ \bottomrule
\end{tabular}
\vspace{-1.5em}
\end{table}

\section{Conclusion}
In summary, we propose a new paradigm for the text-based speech insertion task that enables any-length insertion and full sentence generation. It benefits from decomposing local and global factors and controlling them separately. Specifically, we leverage Retriever \cite{yin2022Retriever} to obtain powerful global factor modeling ability and achieve state-of-the-art speaker similarity. We exploit prosody context information with a prosody smoothing task and achieve satisfactory prosody continuity. We eliminate spectrum over-smoothing with an adversarial training stage. Currently, we have not experimented with our method in highly noisy or reverberant cases, and we make it as our future work.

\bibliographystyle{IEEEtran}

\bibliography{mybib}

\end{document}